\documentstyle[12pt,twoside,a4]{article}

\renewcommand{\theequation}{\thesection.\arabic{equation}}

\newcommand{\rbox}[1]{\vbox{\hrule height.8pt%
                \hbox{\vrule width.8pt\kern5pt
                \vbox{\kern5pt\hbox{#1}\kern5pt}\kern5pt
                \vrule width.8pt}
                \hrule height.8pt}}
\begin{document}
\renewcommand{\thepage}{}

\begin{titlepage}
\title{
\hfill
\parbox{4cm}{\normalsize KUNS-1367 \\HE(TH)~95/17\\
hep-ph/9510426}\\
\vspace{4ex}
Duality of a Supersymmetric Model\\
 with the Pati-Salam group
\vspace{7ex}}

\author{Nobuhiro Maekawa
\thanks{E-mail address: {\tt maekawa@gauge.scphys.kyoto-u.ac.jp}}
and Tomohiko Takahashi\thanks{JSPS Research Fellow. E-mail address:
 {\tt tomo@gauge.scphys.kyoto-u.ac.jp}}
\vspace{3ex}\\
  {\it Department of Physics, Kyoto University}\\
   {\it Kyoto 606-01, Japan}}
\date{}

\maketitle
\vspace{9ex}

\begin{abstract}
\normalsize
\baselineskip=17pt plus 0.2pt minus 0.1pt
Recently one of the authors proposed a dual theory of a
Supersymmetric Standard Model (SSM), in which it is naturally
understood that at least one quark (the top quark) should be heavy, i.e.,
almost the same
order as the weak scale, and 
the supersymmetric Higgs mass parameter $\mu$
can naturally be expected to be small. 
However, the model cannot have Yukawa couplings of the lepton sector.
In this paper,
we examine a dual theory of a Supersymmetric Model with the Pati-Salam
gauge group $SU(4)_{PS}\times SU(2)_L\times SU(2)_R $ with respect to
the gauge group $SU(4)_{PS}$.
In this scenario,  Yukawa
couplings of the lepton sector can be induced.
In this model the Pati-Salam breaking scale
should be around the SUSY breaking scale.
\end{abstract}
\end{titlepage}

\renewcommand{\thepage}{\arabic{page}}
\renewcommand{\theequation}{\arabic{equation}}
\setcounter{page}{1}
\baselineskip=19pt plus 0.2pt minus 0.1pt
\noindent
Recently, it has become clear that certain aspects of four dimensional 
supersymmetric field theories can be analyzed exactly~\cite{duality,
holomorphy,both,DSB}. By using the innovation, it has
been tried to build models in order to solve some phenomenological
problems~\cite{DSB,mike,IzaYana,Stras}. 
One of the most interesting aspects is ``duality''~\cite{duality,both}. 
By using
``duality'', we can infer the low energy effective theory of a strong
coupling gauge theory.
One of the authors suggested that nature may use this ``duality''. 
He discussed a duality of a Supersymmetric Standard 
Model(SSM). But unfortunately, his model does not have Yukawa couplings
of the lepton sector. In this paper, we would like to discuss a duality of
a supersymmetric(SUSY) model with Pati-Salam gauge
group~\cite{PatiSalam} in order to 
obtain all Yukawa couplings.

First we would like to review Seiberg's duality.
Following his discussion~\cite{duality}, we examine $SU(N_C)$ SUSY QCD
with 
$N_F$ flavors of
chiral superfields,
\vskip 0.5cm
\begin{center}
\begin{tabular}{|l|c|c|c|c|c|} \hline \hline
       & $SU(N_C)$ & $SU(N_F)_L$ & $SU(N_F)_R$ & $U(1)_B$ & $U(1)_R$ \\ \hline
$Q^i$    & $N_C$ & $ N_F$ &  1  &   1    & $(N_F-N_C)/N_F$  \\
$\bar Q_j$ & $\bar N_C$ & 1 & $\bar N_F$ & $-1$ & $(N_F-N_C)/N_F$ \\
\hline \hline
\end{tabular}
\end{center} 
\vskip 0.5cm
which has the global symmetry $SU(N_F)_L\times SU(N_F)_R \times U(1)_B
\times U(1)_R$.  
In the following, we would like to take $N_F\geq N_C+2$, though in the
case $N_F\leq N_C+1$ there are a lot of interesting
features~\cite{both,ADS,AKMRV,NSVZ}. 
Seiberg suggests~\cite{duality}
that in the case $N_F\geq N_C+2$ at the low energy scale 
the above theory is 
equivalent to the following $SU(\tilde N_C)$ SUSY QCD theory $(\tilde
N_C=N_F-N_C)$ with $N_F$ flavors of chiral superfields $q_i$ and $\bar 
q^j$ and meson fields $T^i_j$,
\begin{center}
\begin{tabular}{|l|c|c|c|c|c|} \hline \hline
       & $SU(\tilde N_C)$ & $SU(N_F)_L$ & $SU(N_F)_R$ & $U(1)_B$ &
       $U(1)_R$ \\ \hline 
$q_i$    & $\tilde N_C$ & $ \bar N_F$ &  1  &  $N_C/(N_F-N_C)$ & $N_C/N_F$  \\
$\bar q^j$ & $\bar {\tilde N_C}$ & 1 & $ N_F$ & $-N_C/(N_F-N_C)$ & $N_C/N_F$ \\
$T^i_j$ & 1 & $N_F$ & $\bar N_F$ & 0 & $2(N_F-N_C)/N_F$ \\
\hline \hline 
\end{tabular}
\end{center} 
\vskip 0.5cm
\noindent
and with a superpotential
\begin{eqnarray}
W=q_iT^i_j\bar q^j.
\end{eqnarray}
The above two theories satisfy the 't Hooft 
anomaly matching conditions~\cite{tHooft}. 
Moreover Seiberg showed that they are
consistent with the decoupling theorem~\cite{decoupling}. 
Namely, if we introduce a mass 
term only for superfields $Q^{N_F}$ and $\bar Q_{N_F}$
\begin{eqnarray}
W=m Q^{N_F}\bar Q_{N_F}
\end{eqnarray}
in the original theory, the dual
theory has vacuum expectation value(VEVs)
$\left<q\right>=\left<\overline{q}\right>=\sqrt{m}$ 
and $SU(N_f-N_c)$ is broken to $SU(N_f-N_c-1)$, 
which is consistent with
the decoupling of the heavy quark in the original theory.

Second, we would like to review a duality of a SUSY Standard 
Model(SSM)~\cite{mike}. 
We introduce ordinary matter superfields
\begin{eqnarray}
&&Q^i_L=(U_L^i, D_L^i):(3,2)_{1\over 6},\quad U_{Ri}^{c}:(\bar
3,1)_{-{2\over 3}},\quad  
D_{iR}^{c}:(\bar 3,1)_{1\over 3} \nonumber \\
&&L^i=(N_L^i,E_L^i):(1,2)_{-{1\over 2}},\quad
E_{Ri}^{c}:(1,1)_1,\quad\quad i=1,2,3, 
\end{eqnarray}
which transform under the gauge group $SU(3)_{\tilde C}\times
SU(2)_L\times U(1)_Y$. There are no Higgs superfields.
We would like to examine the dual theory of this theory with respect
to the gauge group
$SU(3)_{\tilde C}$. In the following, we neglect the lepton
sector for simplicity. Since $N_F=6$, the dual gauge
group is also $SU(3)_C$ ( $\tilde N_C=N_F-N_C$ ), which we would like 
to assign to the QCD gauge group. 
A subgroup, $SU(2)_L\times U(1)_Y$, of 
the global symmetry group $SU(6)_L\times
SU(6)_R\times U(1)_B\times U(1)_R$ is  gauged.
When we assign $Q=(U_L^1,D_L^1,U_L^2,D_L^2,U_L^3,D_L^3)$ and $\bar
Q=(U_R^{c1},D_R^{c1},U_R^{c2},D_R^{c2},U_R^{c3},D_R^{c3})$, the
$SU(2)_L$ generators are given by
\begin{eqnarray}
I_L^a=I_{L1}^a+I_{L2}^a+I_{L3}^a,\quad a=1,2,3,
\end{eqnarray}
where $I_{Li}^a$ are generators of $SU(2)_{Li}$ symmetries which
rotate $(U_L^i, D_L^i)$,
and the generator of hypercharge $Y$ is given by
\begin{eqnarray}
Y={1\over 6}B-(I_{R1}^3+I_{R2}^3+I_{R3}^3),
\end{eqnarray}
where $I_{Ri}^a$ are generators of $SU(2)_{Ri}$ symmetries which
rotate $(U_{Ri}^c, D_{Ri}^c)$.
In this theory, the global symmetry group is $SU(3)_{QL}\times
SU(3)_{UR}\times SU(3)_{DR}\times U(1)_B\times U(1)_R$.
Then we can write down the quantum numbers of dual fields;
\begin{eqnarray}
&&q_{Li}=(d_{Li}, -u_{Li}):(3, \bar 2)_{1\over 6},\quad u_R^{ci}:(\bar
3,1)_{-{2\over 3}},\quad  
d_R^{ci}:(\bar 3,1)_{1\over 3} \nonumber \\
&&M^i_j:(1,2)_{-{1\over 2}},\quad N^i_j:(1,2)_{1\over 2}
\end{eqnarray}
under the standard gauge group $SU(3)_C\times SU(2)_L\times U(1)_Y$.
Here $M^i_j\sim Q_L^iU_{Rj}^c$ and $N^i_j\sim Q_L^iD_{Rj}^c$ are the
meson fields and we assign 
$q=(d_L^1, -u_L^1, d_L^2, -u_L^2, d_L^3, -u_L^3)$ and $\bar
q=(d_{R}^{c1}, -u_{R}^{c1}, d_{R}^{c2}, -u_{R}^{c2}, d_{R}^{c3}, -u_R^{c3})$. 
It is interesting that the matter contents of both theories are almost
the 
same. The difference is the existence of nine pairs of Higgs
superfields $M^i_j$ and $N^i_j$ 
and their Yukawa terms coupling to ordinary matter superfields,
\begin{eqnarray}
W=-q_L^i N_i^j u_{Rj}^c+q_L^iM_i^jd_{Rj}^c.
\end{eqnarray}

It is interesting that the Yukawa couplings can be expected to be of
order 1 because of the strong dynamics. This means that at least one
quark has a heavy mass, which is almost the order of the weak scale
$v$. It is also interesting that the SUSY mass terms of Higgs
particles are forbidden by the global symmetry  $SU(3)_Q\times
SU(3)_{UR}\times SU(3)_{DR}$. 

Unfortunately, the model has a lot of phenomenological problems.
There is no Yukawa coupling of leptons, there appear Nambu-Goldstone
bosons when the Higgs particles have vacuum expectation values, and
the SUSY mass terms of Higgs particles vanish. Moreover we may give 
up the success of the unification of the three gauge couplings,
because we cannot trace the running of the $SU(3)_{\tilde C}$
coupling. 

In the following, we try to avoid the problem of Yukawa couplings of
leptons and of the Nambu-Goldstone bosons.
Why does not the above model have Yukawa couplings of leptons? This is 
because the leptons have no color charge. Therefore, we can expect that 
if we adopt the Pati-Salam gauge group $SU(4)_{PS}$ as the dual gauge
group, the model has the Yukawa couplings of leptons. 

We consider the dual gauge group of $SU(4)_{PS}$. 
The model which is extended minimally from MSSM has six flavors
because a color triplet and a singlet belong to a quartet of 
$SU(4)_{PS}$. In this case, the dual group is $SU(2)$, and it is
necessary to treat the model differently, since the model satisfies
$N_F=3N_C$ and is no longer in a non-abelian coulomb phase. The
gauge coupling may not become strong. Therefore, we investigate another
possibility of realizing $SU(4)_{PS}$. 
If we
introduce fourth generation, we should impose some unnatural mass
relations in order to suppress the Peskin-Takeuchi's S and T
parameters \cite{PeskinTake}. 
Therefore, we would like to add one vector-like
generation. Since $N_F$ becomes ten, the dual gauge group of the
$SU(4)_{PS}$ becomes $SU(6)$.  
Here we introduce the following superfields:
\begin{eqnarray}
\Psi^i_L:(6,\bar 2,1),\quad \Phi_L :(6,1,\bar 2), 
\quad \Psi_{Ri}^{c}:(\bar 6,1,2),\quad \Phi_R^{c} :(\bar 6, 2, 1),
\quad\quad i=1,2,3,4, 
\end{eqnarray}
under the gauge group
$SU(6)_{HC}\times SU(2)_L\times SU(2)_R$.
Since $N_F=10$ under the gauge group $SU(6)_{HC}$, 
the dual gauge group becomes $SU(4)_{PS}\times SU(2)_L\times SU(2)_R$.
Then the dual fields become
\begin{eqnarray}
\psi^i_L:(4, 2,1),\quad \phi_L :(4,1, 2), 
\quad \psi_{Ri}^{c}:(\bar 4,1,\bar 2),\quad \phi_R^{c} :(\bar 4, \bar 2, 1),
\nonumber\\
T^i_j:(1,\bar 2, 2),\quad M_i^{(1)}:(1,1,1), \quad
M_i^{(3)}:(1,3,1),\\
N^{(1)i}:(1,1,1),\quad N^{(3)i}:(1,1,3),\quad S:(1,2,\bar 2),\nonumber
\end{eqnarray}
which transform under the gauge group
$SU(4)_{PS}\times SU(2)_L\times SU(2)_R$,
and have the superpotential
\begin{eqnarray}
W=\psi_L^iT_i^j\psi_{Rj}^c+\psi_L^i(M_i^{(1)}+M_i^{(3)})\phi_R^c
  +\phi_L(N^{(1)i}+N^{(3)i})\psi_{Ri}^c+\phi_L S \phi_R^c.
\end{eqnarray}
Namely this model has three generations and one vector-like generation with
the Pati-Salam gauge group.
You should notice that the Yukawa couplings of leptons appear. 
If we introduce soft SUSY breaking terms
\begin{eqnarray}
{\cal L}_{SB}^e&=&\sum_{i=1}^4 \left( m_{\Psi L i}^2
  |\Psi_L^i|^2+m_{\Psi R i}^2|\Psi_{Ri}^c|^2\right)+m_{\Phi L}^2
  |\Phi_L|^2+m_{\Phi
    R}^2|\Phi_R^c|^2 \nonumber \\
   && +\left(A_i\Psi_L^i\Phi_R^c+B^i\Phi_L\Psi_{Ri}^c 
   +{1\over 2}\sum_{a=6,2l,2r} \mu_a \lambda_a\lambda_a + h.c.\right),  
\end{eqnarray}
in the original theory, 
all the global symmetries $SU(4)_L\times SU(4)_R\times
\left[U(1)\right]^4$ except $U(1)_{B+L}$ can be broken
explicitly. Therefore, we can avoid the massless Nambu-Goldstone
bosons.

Here we only assume that the above duality can be realized even with
the SUSY breaking terms. 
We can obtain the SUSY breaking terms of the dual theory,
\begin{eqnarray}
{\cal L}_{SB}^e&=&\sum_{i=1}^4 \left( m_{\psi L i}^2
  |\psi_{Li}|^2+m_{\psi R i}^2|\psi_R^{ci}|^2+m_{M1i}^2|M^{(1)}_i|^2
   +m_{M3i}^2|M^{(3)}_i|^2\right.\nonumber\\
   &&\left.+m_{N1i}^2|N^{(1)i}|^2 
   +m_{N3i}^2|N^{(3)i}|^2\right)\nonumber \\
   &&+\sum_{i,j=1}^4 \left(m_{Tij}^2|T^i_j|^2\right)
   +m_{\phi L}^2|\phi_L|^2+m_{\phi
    R}^2|\phi_R^c|^2+m_S^2|S|^2 \\
   &&+\left(\sum_{i=1}^4 \left(A_i\mu N^{(1)i}+B^i\mu M^{(1)}_i\right) 
   +{1\over 2}\sum_{a=4,2l,2r} \mu_a \lambda_a\lambda_a + h.c.\right),
    \nonumber  
\end{eqnarray}
where we treat SUSY breaking parameters perturbatively~\cite{Peskin}
and $\mu$ denotes a typical scale of the dual dynamics. The scale $\mu$
should be larger than the SUSY breaking scale for the perturbation to
be good approximation.\footnote{
Phenomenologically Higgsino masses, which are induced
by the higher order of the perturbation,  should be larger
than the weak scale. In order to realize this situation, the scale
$\mu$ cannot be much larger than the Pati-Salam scale.}
From the above SUSY breaking terms, we can find 
that the scalar fields $ N^{(1)}_i $ and  $M^{(1)}_i$ have vacuum
expectation values(VEVs) of order $\mu$. Therefore, under the
scale $\mu$ we will get the three family model with the Pati-Salam gauge
group. 

In this scenario, however, the Pati-Salam scale should be around the SUSY
breaking scale because of the following two reasons.
The first reason is that there is no vacuum which can break the gauge
symmetry $SU(4)_{PS}\times SU(2)_L\times SU(2)_R$ to the standard
gauge group  $SU(3)_C\times SU(2)_L\times U(1)_Y$ in the flat
direction. 
Namely, the Pati-Salam scale should be the order of the SUSY breaking
scale for the potential problem. The second reason is that the Yukawa
couplings of leptons become too small if the Pati-Salam scale is much
larger than the weak scale. Namely since under the Pati-Salam scale,
``leptons'' in $\psi^i_L$ become massive with $T^4_i$ and ordinary
leptons become three linear combinations of $T^i_j$, 
the Yukawa couplings of the lepton sector become very small.

It seems to be impossible for such a lower Pati-Salam scale to be
consistent with the  
experimental bounds.
From the bound~\cite{data} $\Gamma(K_L^0\rightarrow e\mu
)/\Gamma_{K_L^0}<3.3\times 10^{-11}$, the
Pati-Salam scale is usually estimated to be larger than 1400
TeV~\cite{ValWil}.  
However in the scenario where the $\tau$ lepton is associated with the
down quark, the lower bound of the Pati-Salam scale becomes 13
TeV~\cite{ValWil}, which is not so far from the weak scale as 1400 TeV. 
Therefore, if 
the SUSY breaking scale is of order 10 TeV, we may satisfy the above
experimental constraint. Though such a large SUSY breaking scale is
unnatural, it can suppress flavor changing neutral currents. In future
the signal of 
$B_S^0 \rightarrow \mu e$ may be found~\cite{ValWil}.

Though the structure of quark and lepton mass matrices are too
complicated for us to analyze them, we should comment about the masses
of neutrinos. In this model, two right-handed neutrinos have Majorana masses, 
which are order of the Pati-Salam scale, as a result of their mixing with
gauginos( you should notice that the SUSY 
breaking scale is also of order the Pati-Salam scale ). Therefore,
two left-handed neutrinos become light by seesaw mechanism. However,
since one left-handed neutrino cannot use the seesaw mechanism, the
Dirac masses of the left-handed neutrino should be less than of order
30 MeV, which is the experimental upper bound of the tau neutrino
mass. In this case, other two neutrinos will be lighter than $(10
{\rm MeV})^2/(10 {\rm TeV}) \sim$ 10 eV. 
 

In summary, we examine duality of a SUSY model with the Pati-Salam gauge
group. In this model, the Yukawa couplings of the lepton sector as well as 
of the quark sector can be induced. Since the Pati-Salam scale should be 
around SUSY breaking scale in this model, we should take the SUSY
breaking scale larger than 10 TeV. The signal $B_S^0 \rightarrow \mu
e$ may be found in future experiments.

\section*{Acknowledgments}

We are grateful to the organizers of the 1995 Ontake Summer Institute
and to M. Peskin for a stimulating set of lectures at the institute.
We would like to thank our colleagues for
discussions on ``duality''.
N.M. also thanks M.~Strassler and T.~Yanagida for useful discussions.

\end{document}